%% file: paper.tex
\documentclass[twocolumn,aps,prb]{revtex4-2}

\usepackage{graphicx}
\usepackage[dvipsnames]{xcolor}

\definecolor{bdiv2-0}{HTML}{7ed9e1}
\definecolor{bdiv2-1}{HTML}{58aebf}
\definecolor{bdiv2-2}{HTML}{3f8597}
\definecolor{bdiv2-3}{HTML}{345a68}
\definecolor{bdiv2-4}{HTML}{2b343e}
\definecolor{bdiv2-5}{HTML}{4f2330}
\definecolor{bdiv2-6}{HTML}{91263c}
\definecolor{bdiv2-7}{HTML}{bd494c}
\definecolor{bdiv2-8}{HTML}{de725a}
\definecolor{bdiv2-9}{HTML}{ff9b68}

\definecolor{webblue}{HTML}{2D3092}
\usepackage[colorlinks, urlcolor=webblue, citecolor=webblue, linkcolor=webblue]{hyperref}

\usepackage{xspace}
\usepackage{amsmath}
\usepackage{amsfonts}
\usepackage{amssymb}
\usepackage{mathtools}
\usepackage{braket}
\usepackage{comment}
\usepackage[capitalize]{cleveref}
\usepackage{sidecap}
\usepackage{chemformula}
\usepackage{enumitem}

\newcommand{\vdagger}{\vphantom{\dagger}}

\newcommand{\dd}{\mathrm d}
\newcommand{\bvec}[1]{\boldsymbol{#1}}

\newcommand{\bv}[1]{\boldsymbol{#1}}
\renewcommand{\Im}{\mathrm{Im}}

\input{tikz-settings}

\usepackage{makerobust}
\newcommand{\makeauthor}[2]{\newcommand{#1}[1]{{%
  \protect%
  \color{#2}{%
    \bfseries%
    \begingroup\escapechar=-1\edef\x{\endgroup\string#1}\x:%
  }\itshape{} ##1}}%
  \MakeRobustCommand#1}
\makeauthor{\md}{Plum}
\makeauthor{\lk}{ForestGreen}
\makeauthor{\mk}{orange}
\makeauthor{\ab}{blue}

\newcommand{\prlparagraph}[1]{\textit{#1}---}

\makeatletter
\def\maketitle{
\@author@finish
\title@column\titleblock@produce
\suppressfloats[t]}
\makeatother

\let\oldcite\cite
\renewcommand{\cite}[1]{\if\relax\detokenize{#1}\relax\textbf{\color{red}[?]}\else\oldcite{#1}\fi}

\begin{document}
\title{Pair density wave order from non-symmorphic momentum symmetry}
\author{Matteo Dürrnagel}
\email{matteo.duerrnagel@uni-wuerzburg.de}
\author{C. Alexander Baum}
\author{Michael Klett}
\author{Lennart Klebl}
\email{lennart.klebl@uni-wuerzburg.de}
\author{Ronny Thomale}
\email{ronny.thomale@uni-wuerzburg.de}
\affiliation{Institut für Theoretische Physik und Astrophysik and
Würzburg-Dresden Cluster of Excellence ctd.qmat, Universität Würzburg, 97074
Würzburg, Germany}
\date{\today}

\begin{abstract}
We develop a class of microscopic lattice models in which pair density wave order emerges in the asymptotically exact weak coupling limit. The underlying mechanism is due to non-symmorphic momentum symmetry implied by the models' projective space group representation of electrons. 
Our mechanism suggests moir\'e systems and extended $s$-wave altermagnets as potential hosts for pair density wave order. 
\end{abstract}

\maketitle

\prlparagraph{Introduction}%
Pair density wave (PDW) order, i.e., a superconductor periodically modulated such that its spatial average vanishes, has been a pivotal source of phenomenological inspiration to explain high-$T_c$ copper oxide superconductors~\cite{RevModPhys.87.457}.
This particularly applies to the conception of a PDW mother state as a phenomenological organising principle.
There might be no PDW long range order but significant short range PDW correlations that connect to a series of descendant vestigial orders~\cite{PhysRevX.4.031017}, a motif recently also carried over to alternative mother state theories such as excitonic condensates~\cite{ingham2025vestigialorderexcitonicmother}.
While numerous approximative meanfield PDW ansätze exist aiming at explaining the emergence of experimentally suspected PDW signatures in various strongly correlated electronic scenarios~\cite{Chen-nature2021,doi:10.1126/science.abd4607,Gu-Nature2023,Zhao-Nature2023,PhysRevX.13.031030}, analytically controlled model realizations of PDW order are rare~\cite{annurev:/content/journals/10.1146/annurev-conmatphys-031119-050711}.
This even applies to the weak coupling regime, where at least the interaction strength would not preclude such a finding. The reason is the dominant nesting condition for the $\bv q=0$ pairing susceptibility, generically trumping any other susceptibility channel at finite $\bv q$.
To date, realizing a weak coupling PDW-type state despite this obstacle either necessitates the breaking of time reversal symmetry such as for the FFLO mechanism~\cite{FF,LO}, a fine-tuning of multi-pocket Fermiology~\cite{PhysRevB.107.224516}, a topological flat band tuning such as in~\cite{PhysRevLett.105.215303,PhysRevLett.99.097202} to achieve quantum geometric nesting~\cite{wu2026exactlysolvablepairdensitywave,PhysRevX.14.041004}, or Fermi level tuning to accomplish maximal sublattice interference~\cite{PhysRevB.86.121105,PhysRevB.108.L081117,PhysRevB.107.045122,PhysRevB.110.024501}. 

Non-symmorphic momentum space symmetries (NSMS) embody a recent development of translation symmetries in momentum space that do not correspond to a full reciprocal lattice vector~\cite{Zhang2023g}.
Such symmetries cannot derive from mere crystallography, but rather necessitate a projective space group representation which, upon Fourier transform, then relates to such fractional momentum translation symmetry. 
Emergent NSMS have been recently identified in effective downfolded models of $M$-point twisted moir\'e materials~\cite{calugaru-Nature2025} and bilayer models accomplishing an extended $s$-wave altermagnet~\cite{dürrnagel2026extendedswavealtermagnets}.
Naturally, the question arises which kind of phases might be facilitated by NSMS.
In the case of an extended $s$-wave altermagnet, the anti-unitary symmetry formed by time reversal and a momentum translation by $(\pi,\pi)$ lays the foundation for spin-polarized bands without spin-orbit coupling, and from there a moment-compensated collinear spinful particle-hole condensate without nodes at $\Gamma$, contrasting a conventional altermagnetic spin symmetry group classification~\cite{dürrnagel2026extendedswavealtermagnets}.
Turning to superconducting pairing instabilities, such NSMS is expected to crucially affect the nature of nested pairing, and from there the total momentum of the particle-particle (pp) condensate.

In this Letter, we unfold a microscopic model of weakly coupled electrons featuring a leading pair density wave instability.
Instead of being tied to fine-tuned or approximatively considered interactions, Fermiology, sublattice structure, flat bands, or external fields, this formation mechanism of PDW is exact, requires no fine-tuning, and solely hinges on a projective space group representation of electrons. 
$M$-point twisted moir\'e materials suggest themselves as prime candidates for superconducting order affected by NSMS. Even though the quasi-1D bands of the most promising compound candidates do not immediately suggest any particular propensity to strong Fermi surface instabilities, the specifics of some representatives such as \ch{SnSe2} might bring about unexpectedly high spin fluctuations for unconventional pairing~\cite{klebl2026extendedswavesuperconductivitympoint}.
At the level of fundamental theoretical model building, however, it is desirable to start from a bare discrete lattice model rather than an effective downfolded model that has already been highly processed, such as those models put foward in the context of twisted moir\'e materials.
In order to establish our proposed PDW mechanism, we thus choose a bilayer lattice homogeneously decorated with $\pi$ fluxes between layers.
While such a setup is typically unachievable through magnetic fields, $\pi$ fluxes, being {\it sui generis} time reversal invariant, feature a large variety of synthesis trajectories attainable within material design such as ligand-mediated hopping~\cite{PhysRevResearch.1.032027}, orbital engineering~\cite{schulz-nc2022}, and moir\'e engineering~\cite{eugenio2025tunable, shi2026gvalley, bao2026moire}.
Our model bilayer instills a robust momentum space translation symmetry by half a reciprocal lattice vector, which acts as a custodial symmetry to a resilient perfect nesting condition for the finite-$\bv q$ pairing susceptibility of PDW. From there, within weak coupling at the level of the Bogoliubov de Gennes mean-field equations, we study the asymptotically exact formation of PDW order.

\begin{figure}
    \centering
    \includegraphics[width=\linewidth]{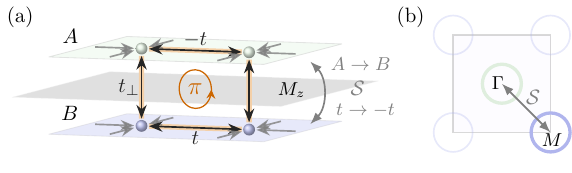}
    \caption{Bilayer lattice model~(a) with non-symmorphic momentum space symmetry (NSMS) $\mathcal S$. NSMS act as layer swap in real space, effectively changing the sign of all intralayer hopping parameters. In momentum space this corresponds to shifting the $\Gamma$ pocket to the $M$ pocket.
    }
    \label{fig:model}
\end{figure}

\prlparagraph{Model}%
We consider a bilayer square lattice [\cref{fig:model}(a)], which comes with hopping parameters of different sign within each of the two layers. The tight-binding Hamiltonian reads
\begin{multline}
    \label{eqn:model}
    H = -t \sum_{\braket{ij},\sigma} \big( c^\dagger_{iA\sigma} c^{\vdagger}_{jA\sigma} - c^\dagger_{iB\sigma} c^{\vdagger}_{jB\sigma} \big) \\
    - t_\perp \sum_{i,\sigma} \big( c^\dagger_{iA\sigma} c^{\vdagger}_{iB\sigma} + c^\dagger_{iB\sigma} c^{\vdagger}_{iA\sigma} \big) \,,
\end{multline}
where $c_{il\sigma}^{\dagger}$ ($c_{il\sigma}^{\phantom{\dagger}}$) creates (annihilates) an electron on site $i$ and layer $l \in \{ A, B \}$  with spin $\sigma \in \{ \uparrow, \downarrow \}$.
Here, $t$ is the in-plane (nearest-neighbor) hopping amplitude alternating in sign between layers and $t_\perp$ denotes the interlayer hopping. (Hopping $t$ is set to 1 in the following.)


The model in \cref{eqn:model} features in-plane $C_{4v}$ point-group symmetry and a nontrivial horizontal mirror $M_z$ that acts projectively [\cref{fig:model}(a),(b)]: the system's symmetry group $\mathcal G$ is characterized by a projective real space representation $\rho$ generated by the projective algebraic relation between real space translations $T_{\bvec a_i}$ illustrated in [\cref{fig:model}(a)] (with $\bvec a_1$, $\bvec a_2$ denoting the real space lattice vectors) and the horizontal mirror $M_z$:
\begin{multline}
    \{ \rho(T_{\bvec a_i}), \rho(M_z) \} = [\rho(T_{\bvec a_i}), \rho(g)] = [\rho(M_z), \rho(g)] = 0 \\
    \forall \, g \in C_{4v}  \,.
    \label{eqn:symmetries}
\end{multline}
The phase factor of the symmetry multiplication, defined by $\rho(g_1) \rho(g_2) = \nu(g_1, g_2) \rho(g_1 g_2)$, is given by $\nu(T_{\bvec a_1 + \bvec a_2},  M_z) = -1$~\cite{Mackey1958}, suggesting form invariance to the magnetic exchange algebra induced by Peierls' phase factors: Layer exchange ($M_z$) acts as vertical hopping, and $M_z T_{\bvec a_i} M_z^{-1} T_{\bvec a_i}^{-1}$ as plaquette hopping around the inserted $\pi$ flux [\cref{fig:model}(a)].
This leads to a non-symmorphic momentum space representation of the projective real space group resulting in the \emph{reciprocal space} wallpaper group $p4g$. The fractional translation elements $\mathcal S\in p4g$ indeed protect the valley degeneracy apparent in \cref{fig:model}(c). Moreover, 
longer range hoppings can readily be added to \cref{eqn:model} as long as they comply with the symmetry group $\mathcal G$.
In several ways, our model is a 2D generalization of the 1D toy model in Ref.~\cite{calugaru-Nature2025}.

\begin{figure}
    \centering
    \includegraphics[width=\linewidth]{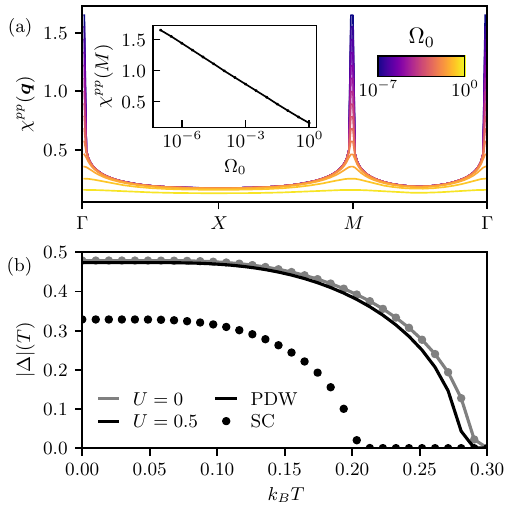}
    \caption{PDW formation in \cref{eqn:model}.
    (a)~Leading eigenvalues of the bare pp susceptibility \cref{eqn:chi_pp} along the high symmetry path for different IR cutoffs $\Omega_0$. The inset shows the logarithmic divergence of the pairing susceptibility at $\bvec q = M$.
    (b)~Temperature dependence of both the SC and PDW gap function with (black) and without (grey) Coulomb repulsion. The gap size $|\Delta|$ is determined by the lowest Matsubara frequency.
    The data is taken at $t_\perp = 0.5$, $\mu = -3$, $\lambda = 2$, and $\omega_0 = 20$; the pp susceptibility are evaluated on a regular $128^2 \times 128^2$ BZ momentum grid.}
    \label{fig:pdw}
\end{figure}

\prlparagraph{Pair density wave order}%
The momentum space translation symmetry by $\bv q=M$ directly implies a perfect pp nesting not only for vanishing momentum transfer ($\bv q=0$), but also with $\bv q=M$. In weak coupling, 
the pp susceptibility dominates:
\begin{equation}
    \chi^{pp}_{nm}(\bv q) = \int\frac{\dd\bv k}{V_\mathrm{BZ}}\, \frac{f(-\varepsilon_n(\bv k + \bv q)) - f(\varepsilon_m(-\bv k))}{\varepsilon_n(\bv k + \bv q) + \varepsilon_m(-\bv k)} \,,
    \label{eqn:chi_pp}
\end{equation}
where $f(x) = 1/(1+e^{\beta x})$ is the Fermi function at inverse temperature $\beta =1/(k_B T)$ and $n,m$ are band indices.
For $\bvec q = 0$, \cref{eqn:chi_pp} generically diverges like $\chi^{pp}(\bvec q = 0) \sim D(\varepsilon_F) \ln(W/\Omega_0)$, i.e., with a Cooper logarithm ($W$ is the system's bandwidth and the infrared (IR) cutoff $\Omega_0 \rightarrow 0$)~\cite{Cooper1956}.
Any arbitrarily small interaction, attractive~\cite{Bardeen1957} or repulsive~\cite{Kohn1965}, will result in an instability of the Fermi surface giving way to SC ($\bv q=0$) order.
In the presence of NSMS such as in \cref{eqn:model}, this conventional wisdom is extended.
\Cref{fig:pdw}(a) depicts the leading eigenvalues of the pp susceptibility for \cref{eqn:model} at different IR cutoffs $\Omega_0$.
We find that the pp susceptibility is peaked both at $\bvec q = 0$ and $\bv q=M$. The two peaks are symmetry-related by the non-symmorphic elements $\mathcal S$ of the NSMS group $p4g$, which enforces a degeneracy between eigenvalues at transfer momenta $\bv q=0, M$.
This can be understood from the spectral structure of \cref{eqn:model} satisfying $\varepsilon_n(\bvec k + M) = \varepsilon_n(\bvec k)$.
Viewed in layer-space instead of band space, the Bloch functions relate the intra-layer structure of $\chi^{pp}(\bv q=0)$ to inter-layer components of $\chi^{pp}(\bv q=M)$ (and vice-versa) via the action of $M_z$, cf.~\cref{eqn:symmetries}.
As a direct result of the perfect $\bvec q = M$ pp nesting, the associated pairing susceptibility diverges logarithmically, in full similarity to $\bvec q = 0$ [\cref{fig:pdw}(a)].

To investigate which of the two degenerate pairing fluctuations will eventually become dominant and result in a phase transition to a symmetry broken phase we add a simple attractive retarded interaction generated by interlayer breathing phonons. After integrating out the phonons we arrive  at a momentum independent effective electron-electron interaction in Matsubara space
\begin{equation}
    V^\mathrm{eph}_{ijkl}(\bvec q, i \nu_n) = \lambda^2 \frac{\omega_0^2}{\nu_n^2 -\omega_0^2} \, (1-\delta_{ij})(1-\delta_{kl}) \,,
\label{eqn:elph}
\end{equation}
where $\nu_n = 2 \pi n/\beta$, $n \in \mathbb N$, $\lambda$ denotes the strength of the electron-phonon coupling, $\omega_0$ is the frequency of the optical breathing phonon, and $i,j,k,l$ are layer indices.
To treat both pairing tendencies ($\bv q\in\{0, M\}$) on equal footing, we solve the frequency dependent gap equation self-consistently on a $\sqrt 2 \, \times\sqrt 2$ supercell, mapping $\bv q=M$ to the reduced zone center.
We make use of spin rotational invariance to decompose the two-particle interaction to singlet (s) and triplet (t) sectors, dropping the spin degree of freedom in the gap equation
\begin{equation}
    \Delta^{s/t}_{ij}(k) = -\int_{k'} V^{s/t}_{ijkl}(k-k') \, F_{kl}(k') \,.
    \label{eqn:gap_equation}
\end{equation}
Here, $k=(\bv k, i\omega_n)$, $\int_k = 1/\beta V_\mathrm{BZ}\sum_{i\omega_n}\int\dd\bv k$, and $F_{ij}(k)$ is the anomalous propagator in the symmetry-broken state (as a matrix in layer/supercell indices $i,j$), defined via the full Bogoliubov-de-Gennes propagator
\begin{equation}
    G_\mathrm{BdG} = \begin{pmatrix} G & F \\ F^\dagger & G^T \end{pmatrix} = \begin{pmatrix}
        i\omega_n - H(\bv k) & \Delta \\
        \Delta^\dagger & i\omega_n + H^T(\bv k)
    \end{pmatrix}^{-1}  \,.
    \label{eqn:gf_bdg}
\end{equation}
Note that the gap functions $\Delta(\bv k,i\omega_n)$ do not obtain a momentum dependence implied by the monentum-independent form of $V^\mathrm{eph}$ (\cref{eqn:elph}).

Solving \cref{eqn:gap_equation} self-consistently for $\Delta$ reveals a conventional intralayer $s$-wave state
\begin{equation}
    \Delta_\mathrm{SC}(\bv k)\propto\braket{c^\dagger_{\bv kA\uparrow}c^\dagger_{\bv -kA\downarrow} + c^\dagger_{\bv kB\uparrow}c^\dagger_{-\bv kB\downarrow} + \mathrm{h.c.}}\,,
\end{equation}
and an interlayer $\bv q=M$ PDW state
\begin{equation}
    \Delta_\mathrm{PDW}(\bv k) \propto \braket{c^\dagger_{\bv k+MA\uparrow}c^\dagger_{-\bv kB\downarrow} + c^\dagger_{\bv k+MB\uparrow}c^\dagger_{-\bv kA\downarrow} + \mathrm{h.c.}}\,,
\end{equation}
residing both in the spin singlet channel. Their gap amplitudes are shown as a function of temperature in \cref{fig:pdw}(b).
Given the commensurate wave vector $\bv q=M$ of the PDW state, its amplitude is uniform and only its phase changes sign between neighboring sites.
This state is an example of a spontaneously formed FF state previously termed $\eta$ pairing~\cite{deBoer1995, Zhai2005, Kaneko2019, Moudgalya2020, Li2020}.

The degeneracy between $\Delta_\mathrm{SC}$ and $\Delta_\mathrm{PDW}$ results from the degenerate pairing susceptibility in \cref{fig:pdw}(a) and the form of the interaction \cref{eqn:elph}. $V^\mathrm{epc}$ couples equally to inter- and intralayer pair fluctuations and therefore equally enhances $\chi^{pp}(\bv q=0)$ and $\chi^{pp}(\bv q=M)$. Even though the states are degenerate, we do not observe mixing down to zero temperature.
Upon inclusion of Hubbard repulsion $H_I = U\sum_i n_{i\uparrow} n_{i\downarrow}$ preserving NSMS, i.e., $[g,H_I] = 0\ \forall g\in\mathcal G$, we find that PDW order is preferred over SC order [\cref{fig:pdw}(b)].
The splitting from nonzero $U$ stems from local penalizing onsite pair formation while leaving bond pairs untouched, hence allowing the PDW state to dominate.
Note that screened Coulomb repulsion is generally expected in electronic systems and conventionally even dominates the phonon-mediated attraction, rendering our scenario generic with regard to the emergence of PDW as the leading instability of a weakly coupled Fermi liquid.
Inclusiong of any in-plane dispersion for the interlayer breathing phonon modes gives the same result as above, given that $U$ remains comparable to the phonon coupling strength.

From the way the PDW is obtained above we find the PDW to be  an asymptotically exact ground state of the system at weak coupling: Other phase transitions of the Fremi liquid (like spin and charge orders) require finite interaction strength in the absence of perfect particle-hole nesting or divergent density of states, so the only instabilities available are those that benefit from the Cooper logarithm.
With NSMS present , SC \emph{and} PDW with commensurate wave vector follow as possible generic weak coupling instabilities, with PDW being selected over SC for nonzero electron-electron repulsion.
In addition to the mean-field results presented in \cref{fig:pdw}(b), we performed electron-phonon functional renormalization group calculations~\cite{Baum2026} to assert that the region of PDW order indeed remains stable for finite interaction strengths $\lambda, U$.

\prlparagraph{Experimental signatures}%
Having established a generic route towards (commensurate) PDW formation, we continue by characterizing the resulting PDW state.
By construction, perfect pp nesting implies a fully gapped Fermi surface in the $\bv q=M$ paired state---in complete analogy to conventional SC.
The absence of typical PDW signatures in other proposals, such as Fermi arcs, raises the question as of how to discriminate PDW against conventional SC in a system with NSMS.
\begin{figure}
    \centering
    \includegraphics[width=\linewidth]{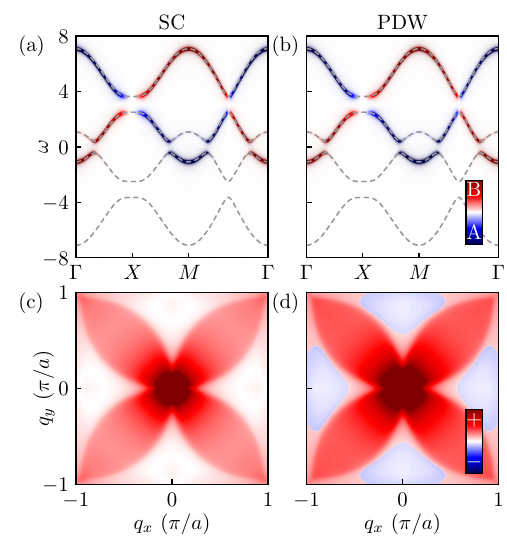}
    \caption{Layer-resolved electronic spectral function $\rho(\bvec k, \omega) = \rho_B(\bvec k, \omega) - \rho_A(\bvec k, \omega)$ in the SC (a) and PDW (b) state at $U = 0$. Grey dashed lines indicate the BdG quasi-particle band structure. 
    There is quantitative agreement between the SC and PDW spectra.
    Phase-referenced Fourier transform $\tilde g_\mathrm{PR}(\bv{q})$ shown for SC~(c) and PDW~(d) for an impurity localized on a single site in the top layer.
    The parameters agree with  \cref{fig:pdw}, i.e., $t_\perp=0.5$ and $\mu=-3$.
    The on-site and bond components of the gap are $\Delta_\mathrm{on-site} = \Delta_\mathrm{bond} \in \{0.5, 0.05\}$ for (a,b) and (c,d) respectively. The QPI spectrum $\tilde g_\mathrm{PR}(\bv{q},E)$ is shown at energy $E = \Delta_\mathrm{eff}/2 \approx 0.0208$.}
    \label{fig:BdG_QPI}
\end{figure}

Resorting to $U = 0$,  the layer-resolved electronic spectral functions for SC and PDW defined as $\rho_i(\bvec k, \omega) = - 1/\pi\,\Im G_{ii}(\bvec k, \omega + i \eta)$ are displayed in \Cref{fig:BdG_QPI} (a) and (b), respectively. 
Intriguingly, the spectra of \cref{fig:BdG_QPI}(a,b) do not only lack qualitative but also quantitative differences when unfolded to the original BZ.
This observation is a direct consequence of NSMS: An interlayer pairing gap between states at $\bvec k + M$ and $-\bvec k$ can be transformed into an intralayer gap with $\bvec k$ and $-\bvec k$, as $M_z$ inverts the layer index of Bloch states at $\bvec k$ and $\bvec k + M$.
Consequently, phase sensitive methods are required to discriminate the two states in an experimental setting. Quasi-particle interference (QPI) has emerged as a pivotal tool to access microscopic details of superconducting order parameters even when in the same symmetry class~\cite{Hanaguri2009, Chi2017, Profe2024,PhysRevLett.130.256001}.
By mapping out the interference pattern of the local density of states (LDOS) around an isolated impurity via the tunneling current between the surface and an atomically thin tip, the sign structure of the gap can be inferred.
The gauge fixing of phase information is achieved through phase referenced Fourier transforms (PRFT) of interference patterns~\cite{Hirschfeld2015, Choubey2014}.
We employ the continuum Green's function method (as implemented in the open source \texttt{calcQPI} code~\cite{Calcqpi1, Calcqpi2}, see documentation for details and implementation) to calculate QPI spectra shown in \cref{fig:BdG_QPI}(c,d).
Naturally, the impurity affects bond- and onsite order parameters in different ways. Even in the present scenario of equal weight in bond- and onsite sectors, the QPI spectra reveal qualitative differences. While the $s$-wave SC spectrum is strictly positive [\cref{fig:BdG_QPI}(c)], the PDW spectrum features a sign change outside the central peak region [\cref{fig:BdG_QPI}(d)].

In addition to phase sensitive probes such as QPI, interacting quantities including dressed charge and spin response functions offer possibilities to discriminate between SC and PDW order.
This point can be connected to the Hubbard interaction $U$ we included to break the degeneracy between SC and PDW:
The NSMS acts in a peculiar fashion on single-particle properties, as these carry only a single momentum index.
Higher-order correlation functions (i.e., two- and more particles) depend on more than one momentum variable and all momenta have to be acted on by $g\in \mathcal G$ simultaneously.
This means that apparent single-particle degeneracies can be broken, allowing for the discrimination of PDW against SC on the mean-field level as well as for unique signatures of the PDW in higher-order correlation functions.


\medskip

\prlparagraph{Note added}%
Upon completion of this manuscript we became aware of a related work reporting an exact pair density wave in moir\'e flat bands. While the approach requires flat band engineering for quantum geometric nesting, it is also based on non-symmorphic momentum symmetry of the chosen twisted bilayer checkerboard lattice~\cite{wu2026exactpairdensitywave}. 

\medskip

\prlparagraph{Acknowledgments}%
We thank T.~Müller for fruitful discussions.
This research was funded by the Deutsche Forschungsgemeinschaft (DFG, German Research Foundation) – Project-ID 258499086 – SFB 1170; through the Würzburg-Dresden Cluster of Excellence on Complexity, Topology, and Dynamics in Quantum Materials (ctd.qmat) – Project-ID 390858490 – EXC 2147; and through the Research Unit QUAST – Project-ID 449872909 – FOR 5249. MD acknowledges support from a PhD scholarship of the Studienstiftung des deutschen Volkes.

\let\oldaddcontentsline\addcontentsline
\renewcommand{\addcontentsline}[3]{}
\bibliography{references.bib}
\let\addcontentsline\oldaddcontentsline

\onecolumngrid
\appendix
\bigskip
\begin{center}
    \bfseries\large End Matter
\end{center}
\medskip
\twocolumngrid

\section{Eliashberg theory}
\label{SM:ET}

To solve the frequency dependent gap equation \cref{eqn:gap_equation} self consistently, we sample the BZ with $200 \times 200$ momentum points and have checked for convergence in the $T \rightarrow 0$ limit with calculations on a $1000 \times 1000$ grid.
To obtain converged results on the Mastubara frequency axis, we employed the \texttt{sparse-ir} library~\cite{wallerberger2023} based on intermediate representation~\cite{shinaoka2017} and sparse sampling~\cite{li2020s}. 
Due to the logarithmic divergence of the pp susceptibility at both $\bvec q = \Gamma$ and $\bvec q = M$ (compare \cref{fig:pdw}) both orders can be treated on equal footing on the MF level.
This is achieved by evaluating \cref{eqn:gap_equation} in the $\sqrt{2} \times \sqrt{2}$ unit cell, such that the layer index $o \in \{A, B\}$ is supplemented by an additional site index $s \in \{1, 2\}$ to form the multi-index $u = (s, o)$. 

\begin{figure}
    \centering
    \includegraphics[width=\linewidth]{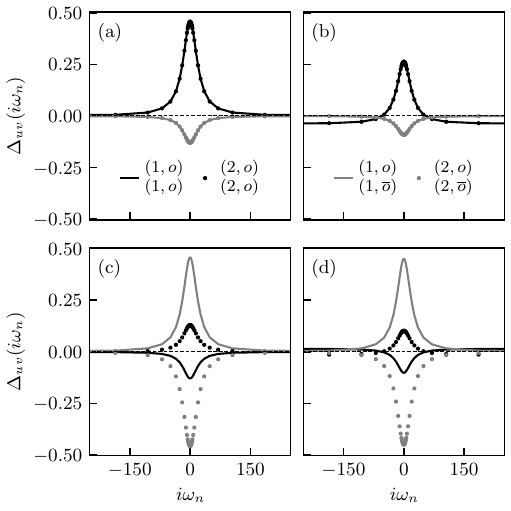}
    \caption{Layer resolved frequency dependent gap functions at $k_B T=0.145$ for the same Hamiltonian parameters as in \cref{fig:pdw}(b).
    Panels (a,b) and (c,d) depict the SC and PDW gap functions, respectively evaluated at $U = 0$ (a,c) and $U=0.5$ (b,d).
    The $\bvec k$ independent gap function is characterized by its multi-index $u = (s, o)$ structure combining both layer $o \in \{A, B\}$ and sublattice $s \in \{1, 2\}$ quantum numbers in the $\sqrt{2} \times \sqrt{2}$ supercell.  
    }
    \label{fig:gap}
\end{figure}

As a direct consequence of the momentum independent bare interaction in the gap equation, the order parameter itself only features a layer and sublattice dependence that is depicted in \cref{fig:gap}.
The PDW (c,d) is characterized by a sign change of the order parameter between the two sublattice sites that is absent in the SC case (a,b).
While in the $U = 0$ case the symmetry $\Delta^{oo}_\text{SC} = \Delta_\text{PDW}^{o \overline o}$ becomes direclty apparent in (a) and (c), the onsite components of both gap functions are heavily suppressed for finite $U$ in (b) and (d) and at the same time exhibit a sign change at $\omega \sim \omega_0$ in accordance with the usual BCS case.
The bond components are only marginally effected by the onsite repulsion instead leading to a more effective gapping of the FS in the PDW case.
Interestingly, no admixture of SC and PDW has been observed in the self-consistent calculations down to $T \rightarrow 0$ independent of the initial conditions.

\begin{figure}
    \centering
    \includegraphics[width=\linewidth]{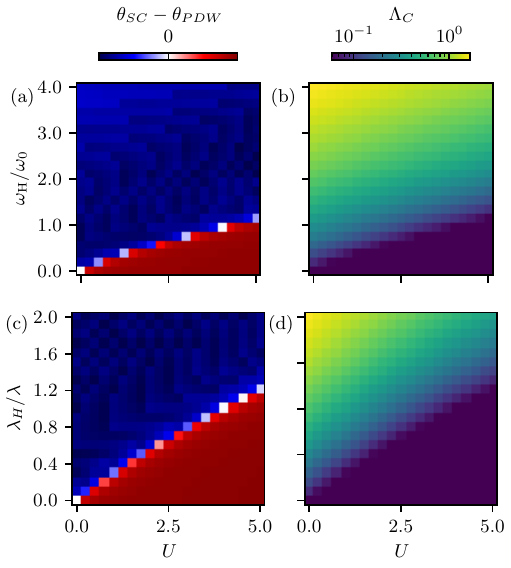}
    \caption{EPFRG phase diagram for the bilayer system with competing phonon modes as a function of phonon frequency (a,b) and EPC (c,d) for varying $U$ at fixed $\lambda = 2$ and $\omega_0 = 20$. The first column depicts the difference between the leading eigenvalues of the pairing vertex at $\Gamma$ ($\theta_{SC}$) and $M$ ($\theta_{PDW}$) and the second column the critical scale of the instability at which the effective two particle vertex diverges.
    }
    \label{fig:phase_diagram}
\end{figure}

\section{Electron phonon functional renormalization group}
\label{SM:FRG}
The EPFRG runs were carried out using the scheme presented in Ref.~\cite{Baum2026}; explicitly with an extension of the open source {divERGe} library~\cite{10.21468/SciPostPhysCodeb.26, 10.21468/SciPostPhysCodeb.26-r0.5} with the TUFRG backend.
For the phase diagram calculations, the bosonic (fermionic) momenta are resolved on a regular $20\times20$ grid in the primitive zone with $10 \times 10$ refinement.
We checked for convergence by comparing with a resolution of $50 \times 50$ ($20 \times 20$) at isolated points.
We truncate the TU expansion on the distance of one primitive lattice vector, since the encountered phases are not expected to display significant NN (or further bond) contributions.
We use {divERGe}'s internal Euler forward integrator to solve the flow equation and set its parameters to:
\begin{itemize}[itemsep=-.3em]
    \newcommand\eulerParam[2]{\item\texttt{#1 = #2;}}
    \eulerParam{eu.Lambda}{100}
    \eulerParam{eu.dLambda\_fac}{0.1}
    \eulerParam{eu.dLambda\_fac\_scale}{30.0}
    \eulerParam{eu.consider\_maxvert\_lambda}{1.e2}
    \eulerParam{eu.lambda\_min}{1.e-3}
\end{itemize}
The flow is terminated when a vertex eigenvalue in the physical channels (charge, magnetic, superconducting; not $P$, $C$, $D$) reaches a value larger than $100$.
The instabilities are determined by diagonalizing the \emph{physical channels}, i.e., charge ($2V^D-V^C$), magnetic ($-V^C$), and superconducting ($V^P$) at every transfer momentum $\bv q$, finding the overall leading eigenvalue $\theta$, and inspecting the corresponding eigenvector(s) for their form factor content.

In addition to the inter-layer breathing phonon mode considered in the main text, we have also investigated the impact of an optical Holstein phonon in each layer. Like the Hubbard-$U$, it breaks the degeneracy between SC and PDW that is also present when considering inter-channel feedback between charge, spin and superconducting fluctuations in the EPFRG by giving rise to an effective retarded attraction
\begin{equation}
    V^\mathrm{H}_{ijkl}(\bvec q, i \nu_n) = \lambda_H^2 \frac{\omega_H^2}{\nu_n^2 -\omega_H^2} \, \delta_{ij}\delta_{kl}
\end{equation}
in the electronic sector.
Mapping out the phase diagram at fixed $\omega_0$ and $\lambda$ of the inter-layer phonon, we obtain a transition between a PDW in the $U$ dominated regime and $s$-wave SC as $\omega_H/\lambda_H$ is successively increased.
As the PDW phase is mostly driven by bond fluctuations, its critical scale is only marginally affected by Hubbard-$U$ and the Holstein phonon mode, both of which act purely onsite on the bare level. The SC phase, on the other hand, is strongly suppressed (enhanced) by the Coulomb repulsion (Holstein phonon) given its onsite nature.

We want to highlight that in the infinitesimal coupling limit, the EPFRG equations reduce the the diagrammatic second order perturbative corrections to the pairing vertex (Kohn-Luttinger diagrams) that are asymptotically exact for $U,\lambda \to 0$~\cite{Raghu2010, Duerrnagel2022}.
Hence, the phase diagram in \cref{fig:phase_diagram} presents an (asymptotically) exact solution to the interacting electron problem in a system with NSMS.

\section{Quasi-particle interference}

\begin{figure}
    \centering
    \includegraphics[width=\linewidth]{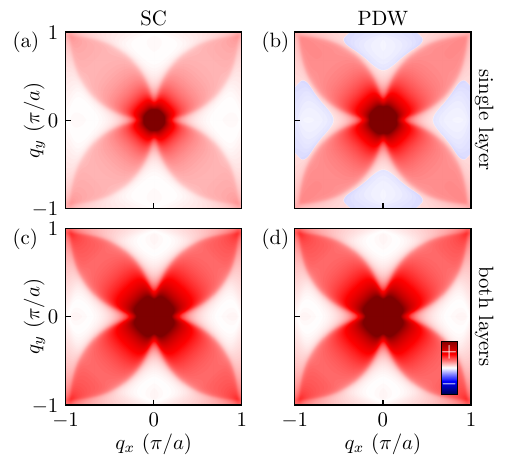}
    \caption{
    Phase-reference Fourier transform for SC and PDW for different measurement probes.
    (a)--(b) is using an impurity that is local to one layer, (c)--(d) has an impurity acting on both layers.
    Here, we use the idealized case of $\delta_\mathrm{on-site} = \delta_\mathrm{bond}$ as defined in \cref{eq:BdG_Hamiltonian}.
    }
    \label{fig:app-QPI_layers}
\end{figure}

QPI as employed here (via the calcqpi package~\cite{Calcqpi1, Calcqpi2}) and detailed in, e.g., Refs.~\cite{Hirschfeld2015, Profe2024} is based on continuum Green's functions.
If we consider the uniform (unperturbed) system, its Green's function reads
\begin{equation}
    \hat G_{0,\sigma}(\bv k, \omega) = \sum_n
    \frac{
    \xi^\dagger_{n\sigma}(\bv k) \xi_{n\sigma}(\bv k)
    }
    {
    \omega - E_{n\sigma} + i\eta
    }\,,
    \label{eq:unperturbed_GF}
\end{equation}
where $\xi_{n\sigma}, E_{n\sigma}$ is the eigenvector/-value of the tight-binding model, $\sigma$ is the spin and $\eta$ a broadening parameter.
An impurity measurement tip is represented by a defect potential that is local to one site $\hat V_{\sigma}$.
We drop spin labels in the following as we discuss nonmagnetic impurity measurements.
The perturbed real-space Green's Function reads
\begin{multline}
    \hat G(\bv{R}, \bv{R'}, \omega) = \hat G_0(\bv{R}-\bv{R'}, \omega)\\
    + \hat G_0(\bv{R}, \omega) \hat T(\omega) \hat G_0(-\bv{R'}, \omega)
    \label{eq:perturbed_GF}
\end{multline}
where $\hat T = \hat V/(\bv{1}-\hat V G_{0}(0,\omega))$ is the $\hat T$-matrix and $\bv{R}$ lattice vectors.
In order to account for nontrivial Wannier centers one may write this in terms of the real-space coordinate
\begin{equation}
    \hat G(\bv{r}, \bv{r'}, \omega) = \sum_{\bv{R},\bv{R'}, \mu, \nu} \hat G(\bv{R}, \bv{R'}, \omega)
    w_{\bv{R},\mu}(\bv{r})
    w_{\bv{R'},\nu}(\bv{r'}) \,,
    \label{eq:RS_GF}
\end{equation}
with $w_{\bv R,\mu}(\bv r)$ the real-space Wannier function on site $\bv R$ and orbital $\mu$. \Cref{eq:RS_GF} implies the continuum local density of states (cLDOS):
\begin{equation}
    \rho(\bv{r},\omega)=-\frac1\pi\Im{\hat G(\bv{r}, \bv{r}, \omega)} \sim g(\bv{r},\omega)\,,
    \label{eq:cLDOS}
\end{equation}
where $g$ is the local differential conductance that is assumed proportional to the cLDOS $\rho$ in a spatially independent manner.
We obtain the QPI signal from phase-referenced Fourier transformation (PR-FFT):
\begin{equation}
    \tilde g_\mathrm{PR}(\bv{q},\omega) = \tilde g(\bv{q},\omega) \left(\frac{\tilde g(\bv{q},\omega_0)}{|\tilde g(\bv{q},\omega_0)|}\right)^{-1} \,,
    \label{eq:PR_FFT}
\end{equation}
where $\tilde g(\bv{q},\omega)$ is the Fourier transform of $g(\bv{r},\omega)$.
The second term in \cref{eq:PR_FFT} effectively divides out a global phase factor in $\tilde g(\bv{q},\omega)$ and allows for the comparison of relative phases between states at energies $\omega$ and $\omega_0$.
One typically phase-references the FFT at $\omega_0 = -\omega$ such that a positive (negative) sign of the interference pattern is related to in-phase (out-of-phase) interference between states at the two energies.

In order to compare SC and PDW we write the Bogoliubov-de-Gennes Hamiltonian in the extended unit cell in $\mathrm{layer}\times \mathrm{sublattice} \times \mathrm{Nambu}$ space:
\begin{align}
    H_\mathrm{BdG}^{\mathrm{SC}/\mathrm{PDW}}(\bv{k})
    &=
    \nu_z H(\bv{k}) + \nu_x \Delta_{\mathrm{SC}/\mathrm{PDW}},
    \\
    \Delta_{\mathrm{SC}/\mathrm{PDW}}
    &=
    (\delta_\mathrm{on-site} \rho_0 + \delta_\mathrm{bond} \rho_x)
    \otimes \begin{cases}
       \tau_0 & \mathrm{SC}\\
       \tau_z & \mathrm{PDW}
    \end{cases},
    \label{eq:BdG_Hamiltonian}
\end{align}
where $H(\bv{k})$ is the tight-binding Hamiltonian in momentum space and $\nu_i$, $\rho_i$ and $\tau_i$ are Pauli matrices acting on Nambu-, layer-, and sublattice-space, respectively.

We observe that SC and PDW differ only in their sublattice dependence, which can be seen from considering the intertwining map
\begin{equation}
    W = \bv{1}_e \oplus (\rho_x \tau_z)_h \,,
    \label{eq:intertwiner_SC_PDW}
\end{equation}
which only exchanges layers in the hole sector in a sublattice dependent manner.
$W$ acts on the BdG Hamiltonian as $H_\mathrm{BdG}^{\mathrm{SC}} = WH_\mathrm{BdG}^{\mathrm{PDW}}W$, i.e., it effectively exchanges SC and PDW states if $\delta_\mathrm{on-site} = \delta_\mathrm{bond}$.
We note that $W$ is not a symmetry of the full Hamiltonian since it only commutes with the kinetic, but not with the interaction part for $U\neq 0$.
This is also why the SC and PDW states are physically distinct and can be distinguished in \cref{fig:pdw} for $U\neq 0$.
It follows that the SC and PDW impurity measurements are equal iff $\mathcal{V} = \hat V \nu_z \otimes \Pi$ commutes with the horizontal mirror $M_z$, where $\Pi$ is the action of the impurity potential in layer and sublattice space.

In \cref{fig:app-QPI_layers} we compare two different probes that both act only on one sublattice position but one acting on a single layer (a,b) and one acting on both layers equally (c,d).
We write
\begin{equation}
    \Pi_{\mathrm{single}/\mathrm{double}} = 
    \begin{pmatrix}
        1 & 0\\
        0 & 0
    \end{pmatrix}_{s}
    \otimes
    \begin{pmatrix}
        1 & 0\\
        0 & 0/1
    \end{pmatrix}_{l}\,,
\end{equation}
where we distinguish layer (l) and sublattice (s) sectors.
Clearly, the double-layer potential commutes with $M_z$ but the single-layer potential cannot. 
This is seen in \cref{fig:app-QPI_layers}(c,d) which exactly match, while (a,b) differ in the PDW response, which has pronounced negative-phase spots.
Intuitively, these spots arise from $\bv{q}$ connecting regions dominated by opposite sublattice character, which have opposite sign for the PDW but not the SC.
Lastly, we note that for $\delta_{\mathrm{on-site}} \neq \delta_{\mathrm{bond}}$, the double-layer measurements of PDW and SC differ quantitatively but not qualitatively, in that no negative-sign lobe appears for the PDW state.
Thus, even in this more general case, only a layer-sensitive measurement can qualitatively distinguish the two states.

\end{document}

%% file: tikz-settings.tex
\usepackage{tikz}
\colorlet{c-v-pppp}{bdiv2-1!70!bdiv2-0}
\colorlet{c-v-pmpm}{bdiv2-4!10!bdiv2-3}
\colorlet{c-v-pmmp}{bdiv2-5!10!bdiv2-6}
\colorlet{c-v-ppmm}{bdiv2-8!70!bdiv2-9}
\definecolor{diffgray}{RGB}{180,180,180}

\usetikzlibrary{arrows, positioning, calc, shapes, decorations.markings,
decorations.pathreplacing, math, snakes, arrows.meta, angles, hobby}
\tikzset{pppp/.style={decorate,decoration={snake,amplitude=0.6mm,segment
length=2.3mm,pre length=0mm,post length=0mm},c-v-pppp, thick}}
\tikzset{pmpm/.style={decorate,decoration={coil,amplitude=1mm,segment
length=1.6mm,pre length=0mm,post length=0mm},c-v-pmpm, thick}}
\tikzset{pmmp/.style={dashed, c-v-pmmp, very thick}}
\tikzset{ppmm/.style={decorate,decoration={zigzag,amplitude=0.6mm,segment
length=3.3mm,pre length=0mm,post length=0mm},c-v-ppmm, thick}}
\tikzset{mf/.style={double distance = .2mm, diffgray, thick}}
\tikzset{midarrow/.style={decoration={markings,mark=at position 0.5 with
{\arrow[xshift=2.5pt]{Latex[length=4pt,#1]}}},postaction={decorate}}}
\tikzset{vp/.style={midway,outer sep=2.5pt,node contents={\tiny$+$}}}
\tikzset{vm/.style={midway,outer sep=2.5pt,node contents={\tiny$-$}}}